\documentclass{IET}

\renewcommand{\baselinestretch}{2}

\usepackage{amsmath,amssymb,amscd}
\usepackage{graphicx,cite}
\usepackage{algorithm}
\usepackage{algpseudocode}
\usepackage{subfigure}

\newcommand{\m}{\text{m}}
\newcommand{\opt}{\text{opt}}

\begin{document}
\renewcommand{\baselinestretch}{1}

\title{Mobile Cloud Computing with a UAV-Mounted Cloudlet: Optimal Bit Allocation for Communication and Computation}

\author[1]{Seongah Jeong}
\affil{School of Engineering and Applied Sciences (SEAS), Harvard university, 29 Oxford street, Cambridge, MA 02138, USA}

\author[2]{Osvaldo Simeone}
\affil{Department of Electrical \& Computer Engineering, New Jersey Institute of Technology (NJIT),
Newark, NJ 07102, USA}

\author[3,*]{Joonhyuk Kang}
\affil{Department of Electrical Engineering, Korea Advanced Institute of Science and Technology
(KAIST), 291, Daehak-ro, Yuseong-gu, Daejeon 34141, Korea}
\affil[*]{jhkang@ee.kaist.ac.kr}

\abstract{Mobile cloud computing relieves the tension between compute-intensive mobile applications and battery-constrained mobile devices by enabling the offloading of computing tasks from mobiles to a remote processors. This paper considers a mobile cloud computing scenario in which the ``cloudlet'' processor that provides offloading opportunities to mobile devices is mounted on unmanned aerial vehicles (UAVs) to enhance coverage. Focusing on a slotted communication system with frequency division multiplexing between mobile and UAV, the joint optimization of the number of input bits transmitted in the uplink by the mobile to the UAV, the number of input bits processed by the cloudlet at the UAV, and the number of output bits returned by the cloudlet to the mobile in the downlink in each slot is carried out by means of dual decomposition under maximum latency constraints with the aim of minimizing the mobile energy consumption. Numerical results reveal the critical importance of an optimized bit allocation in order to enable significant energy savings as compared to local mobile execution for stringent latency constraints.}

\maketitle
\section{Introduction}\label{sec:intro}  
Mobile cloud computing enables the offloading of compute-intensive applications, such as speech or image processing, from mobile devices to a remote processor, with the aim of reducing mobile energy consumption (see, e.g., [1]). The remote processor typically resides in the cloud, and it is accessed by the mobile by means of wireless transmission to a nearby cellular base station, as well as a backhaul connection between base station and cloud. In order to reduce the latency associated with backhaul transmission, an alternative solution has been proposed whereby the remote processor is hosted at a ``cloudlet'', e.g., a PC, that is directly connected to a base station or access point [2].  

For scenarios with limited, or no, existing infrastructure of base stations, recent work has put forth the idea that coverage may be guaranteed by means of moving relays or base stations mounted on unmanned aerial vehicles (UAVs) [3]-[19]. Examples include developing countries or rural environments, as well as in scenarios involving disaster response, emergency relief and military operation. As proposed in [6], UAVs can hence also be used as hosts for cloudlet processors. For instance, thanks to offloading to moving UAVs, battery-limited mobile devices can run computation-intensive application such as for object recognition in emergency relief deployments. The limited coverage and mobility of the energy-constrained UAVs pose new challenge to the design of UAV-based wireless communications systems, as we review in the following. 

\subsection{Related Works}\label{sec:relate}
\textit{UAV as a relay}: In [8]-[12], a UAV-enabled mobile relaying system is studied where the role of the UAV is to act as a \textit{relay} for communication between wireless devices. The problem of jointly optimizing the power allocation at source and moving relay, as well as the relay's trajectory, is tackled in [8] assuming a decode-store-and-forward scheme with the aim of maximizing the throughput under constraints on the relay's speed. To address the problem, an iterative algorithm is proposed to alternatively optimize the power allocation and relay's trajectory. In [9], [10], the problem of efficient data delivery in sparse mobile \textit{ad hoc} or \textit{sensor} networks is studied, where a set of moving relays between pairs of sources and destinations is employed. The authors in [11] study the deployment of UAVs acting as relays between ground terminals and a network base station so as to provide uplink transmission coverage for ground-to-UAV communication. The problem of optimizing the UAV heading angle is tackled with the goal of maximizing the sum-rate under individual minimal rate constraints. Reference [12] proposes a resource allocation optimization mechanism to minimize the mean packet transmission delay in three-dimensional cellular network with multiple-layer UAVs, where the packets from the ground terminals need to be transmitted via several UAV relays to reach the base stations due to the limited transmission range.  

\textit{UAV as a flying base station}: In [13]-[19], wireless communication systems are explored where the role of the UAV is to act as a \textit{flying base station} for ground devices. In [13], a scheduling and resource allocation framework is developed for energy-efficient \textit{machine-to-machine} communications with UAVs, where multiple UAVs provide uplink transmission to collect the data from the heads of the clusters consisting of a number of machine-type devices. The authors in [14] investigate the optimal trajectory and deployment of multiple UAVs to enable reliable uplink communications for ground Internet of Things devices with a minimum energy consumption. References [15] and [16], instead, study the optimal deployment of multiple UAVs acting as flying base stations in the downlink scenario. In particular, the optimal altitudes for the UAVs are addressed with the aim of minimizing the required downlink transmit power for covering a target area in [15]. In contrast, in [16], the UAV's locations and the boundaries of their coverage areas are optimized to minimize the total UAV's downlink transmit power under minimum users' rate requirements. The authors in [17] analyze the downlink coverage and rate performance for static and mobile UAV. Also, the polynomial-time algorithm with successive UAV deployment is proposed in [18] to minimize the number of UAVs needed to provide wireless coverage of a group of ground devices. A point-to-point communication link between the UAV and a ground user is investigated in [19] with the goal of optimizing the UAV's trajectory under a UAV's energy consumption model that accounts for the impact of the UAV's velocity and acceleration.              

\subsection{Problem Statement and Main Contributions}\label{sec:contri}
In this work, we explore the use of a UAV as a moving cloudlet to provide mobile cloud computing opportunities to mobile devices [6]. The main goal is the optimization of the bit allocation for uplink/downlink communication and computing at the cloudlet as a function of the UAV's trajectory, with the aim of minimizing the mobile energy consumption. 

To elaborate, a mobile cloud computing system is considered that consists of a static mobile device and a UAV-mounted cloudlet as illustrated in Fig. \ref{fig:sys}, where the UAV's trajectory is predetermined. This corresponds to the practical scenario where the UAV trajectory is optimized in a preliminary step as a function of the UAV's energy budget, launching/landing locations and pre- and post-mission flying paths, as well as in light of other tasks that the UAV may be carrying out [8], [9], [19]. Use cases for UAV-based edge computing include the support of rescue or military operations via image or video recognition software run on mobile devices for the assessment of the status of victims, enemies, or hazardous terrain and structures. In such cases, such as in earthquake disaster scenarios [9], UAVs equipped with CPU, large storage and short-range radios can be used to provide offloading opportunities, while at the same time also gathering and carrying data among disconnected areas and communicating with victims and rescuers. It is assumed that communication between mobile and UAV takes place by means of frequency division duplex (FDD). Offloading requires communication of the input data for the application in the uplink from the mobile to the UAV, computing at the UAV-mounted cloudlet, and downlink transmission of outcome of the application from UAV to mobile. The problem of optimizing the bit allocation for uplink and downlink communication and computing at the UAV is formulated as a function of the cloudlet's trajectory, and an optimal bit allocation based on dual decomposition is proposed. 

In the rest of the paper, after introducing the system model in Section \ref{sec:sys}, we formulate and solve the mentioned minimization problem in Section \ref{sec:opt}. Numerical results and concluding remarks are presented in Section \ref{sec:num} and Section \ref{sec:con}, respectively.    

\section{System Model}\label{sec:sys}
\subsection{Set-up}\label{sec:setup}
\begin{figure}[t]
\begin{center}
\includegraphics[width=12cm]{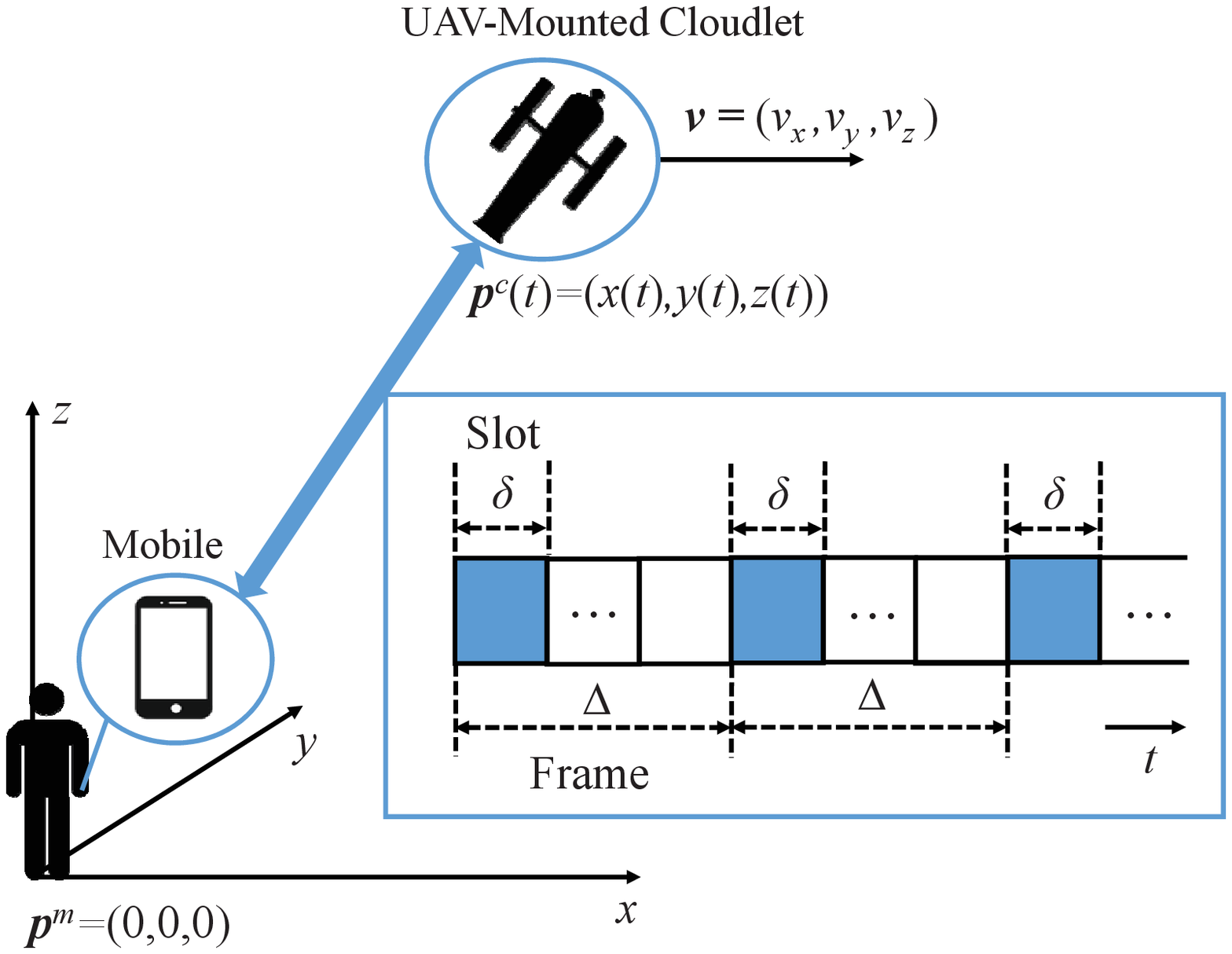}
\caption{Illustration of the considered mobile cloud computing system with a UAV-mounted cloudlet.} \label{fig:sys}
\end{center}
\end{figure}

As illustrated in Fig. \ref{fig:sys}, in this work, a mobile cloud computing system based on a UAV-mounted cloudlet is considered. The focus is on the optimization of the offloading of an application from a given mobile device to the moving cloudlet. Offloading requires communication of the input data for the application to be run at the cloudlet in the uplink from the mobile to the UAV; computing at the UAV-mounted cloudlet; and downlink transmission of outcome of the computation at the cloudlet from UAV to mobile. The mobile application is characterized by the number $L$ of input bits, the number $C$ of CPU cycles per input bit needed for computing, and the number $\kappa$ of output bits produced by computing per input bit produced by the execution of the application.

To describe the system in mathematical terms, a three-dimensional Cartesian coordinate system is considered as illustrated in Fig. \ref{fig:sys}, with all dimensions being measured in meters, where the mobile device is located at $\pmb{p}^\m=(0,0,0)$ and the UAV moves along a trajectory $\pmb{p}^\text{c}(t)=(x(t),y(t),z(t))$, for $t \ge 0$. The UAV's trajectory is assumed to be fixed and known, which depends on its energy budget, landing/launching locations and the pre- and post-mission flying paths [8], [9], [19]. Time is partitioned into frames of duration $\Delta$ seconds, in which the mobile is allocated transmission slots of duration $\delta < \Delta$ for transmission or reception (see Fig. \ref{fig:sys}). The slot duration $\delta$ is chosen to be sufficiently small in order for the UAV's location to be approximately constant within each slot. For the purpose of analysis, the UAV's trajectory $\pmb{p}_c(t)$ can hence be sampled as $\pmb{p}^{\text{c}}_n=(x_n, y_n, z_n) \triangleq \pmb{p}^{\text{c}}(n\Delta)$, where $\pmb{p}^{\text{c}}_n$ is the position of the UAV in the $n$th time slot. 

As in [8], [18], [19], the communication channel between the mobile device and UAV is assumed to be dominated by the line-of-sight component, and that the Doppler effect due to the cloudlet's mobility is perfectly compensated by the receivers. Moreover, FDD with equal channel bandwidth $B$ is assumed to be allocated for uplink and downlink. Accordingly, at slot $n$, the path loss between mobile device to cloudlet is given by
\begin{equation}\label{eq:pl}
h_n = \frac{h_0}{\|\pmb{p}^{\text{c}}_n\|^2}=\frac{h_0}{x^2_n+y^2_n+z^2_n},
\end{equation}   
where $h_0$ represents the received power at the reference distance $d_0=1$ m for a transmission power of $1$ W; and $\|\pmb{p}^{\text{c}}_n\|=\sqrt{x^2_n+y^2_n+z^2_n}$ represents the distance between the mobile device and the UAV at slot $n$. The channel noise is assumed to be additive white Gaussian with zero mean and power spectral density $N_0$ [W/Hz]. 

In this work, we focus on the UAV's energy budget required for communication and computing in the offloading procedure with a predetermined UAV's trajectory. In fact, the energy consumption of the UAV for flying is a constant that depends on the trajectory via the UAV's velocity [20], [21] as well as acceleration [19]. The energy consumption model for computation is first reviewed following [22]-[24]. 
     
\subsection{Computation Energy Model}\label{sec:CompM}
If the frequencies at which the CPUs of the mobile device and cloudlet are operated are given by $f^\m$ and $f^{\text{c}}$, respectively, following [22]-[24], the energy consumptions due to computation of an $l$-bit input are given as 
\begin{equation}\label{eq:E_CP}
E^d(l, f^d) = C\gamma^d (f^d)^2 l,
\end{equation}
where $d=\m$ for the mobile and $d=\text{c}$ for the cloudlet. In (\ref{eq:E_CP}), $\gamma^d$ is the effective switched capacitance of the corresponding device, which is determined by its chip architecture. The model (\ref{eq:E_CP}) indicates that the energy per bit is proportional to the square of the CPU frequency $f^d$. This can be justified by the fact that, when the dynamic power dominant among the CPU power is considered [22], [23], the energy per operation is proportional to the square of the voltage supply $V$ to the chip in CMOS circuits. Moreover, it has been observed that, at the low CPU voltage limits, the frequency $f^d$ of the chip is approximately linear proportional to the voltage supply $V$, which yields the computation energy model (\ref{eq:E_CP}) [24].    
\subsection{Communication Energy Model}\label{sec:CommM}
The energy required to transmit $L^d$ bits in the uplink ($d=\m \to \text{c}$) and in the downlink ($d=\text{c} \to \m$), respectively, within a time slot of duration $\delta$, with a path loss $h$, can be computed based on standard information-theoretic arguments [25] as
\begin{equation}\label{eq:E_CM}
E^d(L^d)=\left(2^{\frac{L^d}{B\delta}}-1\right)\frac{N_0 B \delta}{h}.
\end{equation}
The model (\ref{eq:E_CM}) follows since by equating the number of bits $L^d$ to the maximum number of bits that can be transmitted in a time slot of duration $\delta$, that is [25] 
\begin{equation}
B\delta\log_2\left(1+ \frac{E^d(L^d)h}{N_0 B \delta}\right) = L^d.
\end{equation} 
\section{Optimal Bit Allocation}\label{sec:opt}
In this section, the optimal bit allocation for transmission and computing is studied under a maximum latency constraint of $T$ seconds or, equivalently, $N$ frames with $T=N\Delta$. The energy consumption under mobile execution is first computed in Section \ref{sec:ME} for reference, and then we study the optimization of the offloading process for cloudlet execution in Section \ref{sec:CE}.
\subsection{Energy Consumption for Mobile Execution}\label{sec:ME}
Here, the energy consumption needed to run the application at the mobile is briefly considered for reference. In this case, the mobile device needs to process the $L$-bit input data within $T$ seconds. To this end, the CPU frequency must be selected as  
\begin{equation}\label{eq:optf_ME}
f^{\m} = \frac{CL}{T}, 
\end{equation} 
so that the number of processed bits $f^{\m}T$ equals $CL$. Plugging (\ref{eq:optf_ME}) into (\ref{eq:E_CP}) yields the energy [22]-[24]
\begin{equation}\label{eq:optE_ME_tot} 
E^{\m} \triangleq E^{\m}(L, f^{\m}) = \frac{\gamma^\m C^3}{T^2}L^3.
\end{equation}
\subsection{Optimal Bit Allocation for Cloudlet Execution}\label{sec:CE}
In this section, offloading via cloudlet execution is studied. The time slot of each frame is assumed to be allocated to the given mobile (see Fig. \ref{fig:sys}) used for communication in both the uplink and downlink due to FDD, as well as for executing the application of the mobile device at the cloudlet. We emphasize that this assumption accounts for the fact that the cloudlet generally serves other mobiles in the same frame. To elaborate, for any slot of the $n$th frame, henceforth referred to as the $n$th slot, we define the number of input bits transmitted in the uplink from the mobile device to cloudlet as $L_n^{\m \to \text{c}}$, the number of bits processed at the cloudlet as $l_n^{\text{c}}$, and the number of bits transmission in the downlink from cloudlet to mobile device as $L_n^{\text{c} \to \m}$. Furthermore, the frequency, at which the cloudlet CPU is operated at the $n$th slot, is denoted as $f_n^{\text{c}}$. 

At the first slot, $n=1$, the mobile device transmits $L_1^{\m \to \text{c}}$ bits to the cloudlet in the uplink, without computing or downlink transmission, i.e., $l_1^{\text{c}}=L_1^{\text{c} \to \m}=0$. At the next slot, $n=2$, $L_2^{\m \to \text{c}}$ bits are transmitted in the uplink and the cloudlet computes $l_2^{\text{c}} \le L_1^{\m \to \text{c}}$ bits with the CPU frequency $f_2^{\text{c}}$ without downlink transmission, i.e., $L_2^{\text{c} \to \m}=0$. At the third slot, $n=3$, while $L_3^{\m \to \text{c}}$ bits are transmitted from mobile device and $l_3^{\text{c}} \le L_1^{\m \to \text{c}}+L_2^{\m \to \text{c}} - l_2^{\text{c}}$ bits are computed at the cloudlet with CPU frequency $f_3^{\text{c}}$, the cloudlet transmits $L_3^{\text{c} \to \m}$ bits in the downlink. Given that $l$ bits yield $\kappa l$ bits at the output, we have the constraint $L_3^{\text{c}\to \m} \le \kappa l_2^{\text{c}}$. The procedure is continued until the $N$th frame under the constraint that all input bits are transmitted and processed, that is, $\sum_{n=1}^{N-2} L_n^{\m \to \text{c}}=L$ and $\sum_{n=1}^{N-2} l_{n+1}^{\text{c}}=L$, and all the output bits are retransmitted, i.e., $\sum_{n=1}^{N-2} L_{n+2}^{\text{c} \to \m}=\kappa L$. The CPU frequency at slot $n$ is selected so as to guarantee the processing $l_n^{\text{c}}$ bits within a time slot as
\begin{equation}\label{eq:optf_CE}
f_n^{\text{c}}=\frac{Cl_n^{\text{c}}}{\delta}, 
\end{equation}   
yielding the computation energy consumption at the $n$th slot as a function only of $l_n^{\text{c}}$ as follows: 
\begin{equation}\label{eq:optEk_CE}
E^{\text{c}}(l_n^{\text{c}}) \triangleq E^{\text{c}}(l_n^{\text{c}}, f_n^{\text{c}})= \frac{\gamma^\text{c}C^3}{\delta^2}(l_n^{\text{c}})^3.
\end{equation}    

The optimal bit allocation amounts to the selection of the bit sequences $\{L_n^{\m \to \text{c}}\}_{n=1}^{N-2}$, $\{l_n^{\text{c}}\}_{n=2}^{N-1}$ and $\{L_n^{\text{c} \to \m}\}_{n=3}^{N}$ for communication and computing with the aim of minimizing the mobile energy consumption while satisfying the latency constraint and an energy constraint at the cloudlet. The problem is formulated as follows:
\begin{subequations}\label{eq:CEopt}
\begin{eqnarray}
&& \hspace{-1.5cm} \underset{\{L_n^{\m \to \text{c}} \ge 0\}, \{l_n^{\text{c}} \ge 0\}, \{L_n^{\text{c} \to \m} \ge 0\}}{\text{minimize}} \hspace{0.3cm} \sum_{n=1}^{N-2}  E^{\m \to \text{c}}(L_n^{\m \to \text{c}}) \label{eq:obj}\\
&& \hspace{-0.8cm}  \text{s.t.} \hspace{0.3cm}  \sum_{n=1}^{N-2}  E^{\text{c}}(l_{n+1}^{\text{c}}) + E^{\text{c} \to \m}(L_{n+2}^{\text{c} \to \m}) \le E^{\text{c}}_0\label{eq:lifetime}\\
&& \hspace{0cm}  \sum_{i=1}^{n}l_{i+1}^{\text{c}} \le \sum_{i=1}^{n}L_{i}^{\m \to \text{c}},\hspace{0.1cm} \text{for} \hspace{0.1cm} n=1, \dots, N-2 \label{eq:ineq_mc}\\
&& \hspace{0cm}  \sum_{i=1}^{n}L_{i+2}^{\text{c} \to \m} \le \kappa \sum_{i=1}^{n}  l_{i+1}^{\text{c}}, \hspace{0.1cm} \text{for} \hspace{0.1cm} n=1, \dots, N-2 \label{eq:ineq_cm}\\
&& \hspace{0cm} \sum_{n=1}^{N-2} L_n^{\m \to \text{c}}=L \label{eq:eq_Lmc}\\
&& \hspace{0cm} \sum_{n=1}^{N-2} l_{n+1}^{\text{c}}=L  \label{eq:eq_lc}\\
&& \hspace{0cm} \sum_{n=1}^{N-2} L_{n+2}^{\text{c} \to \m}=\kappa L, \label{eq:eq_Lcm}
\end{eqnarray}
\end{subequations}  
where $E^{\m \to \text{c}}(L_n^{\m \to \text{c}})$ and $E^{\text{c} \to \m}(L_n^{\text{c} \to \m})$ are defined as (\ref{eq:E_CM}) with path loss $h_n$ at each slot $n$ in (\ref{eq:pl}); and $E^{\text{c}}_0$ in (\ref{eq:lifetime}) represents the cloudlet energy budget allocated to the given user for the communication and computing. In problem (\ref{eq:CEopt}), the inequality constraints (\ref{eq:ineq_mc}) enforces that the number of input bits computed at the $n$th slot by the cloudlet be no larger than the number of bits received by the cloudlet in the uplink in the previous $n-1$ slots, for $n=2, \dots, N-1$. Constraint (\ref{eq:ineq_cm}) ensures that the number of bits transmitted from the cloudlet in the downlink at the $n$th slot is no larger than the number of bits available at the cloudlet upon computing in the previous $n-1$ slots, for $n=3, \dots, N$. Finally, the equality constraints (\ref{eq:eq_Lmc}) - (\ref{eq:eq_Lcm}) guarantee that the input bits given at the mobile device are completely processed via offloading within the latency constraint of $N$ frames, or $T$ seconds.

Problem (\ref{eq:CEopt}) is convex. In fact, the objective function (\ref{eq:obj}) is the sum of convex exponential functions; the constraint (\ref{eq:lifetime}) is the sum of convex exponential functions and cubic functions defined in the nonnegative domain; and the constraints (\ref{eq:ineq_mc}) - (\ref{eq:eq_Lcm}) are linear. Accordingly, the problem (\ref{eq:CEopt}) can be numerically solved by standard convex optimization techniques. Instead of relying on a generic solver, here we propose a bit allocation approach based on dual decomposition [26]. To this end, the Lagrangian dual variables $\mu \ge 0$, $a_n \ge 0$ and $b_n \ge 0$ for $n=1, \dots, N-2$ are introduced corresponding to the constraints (\ref{eq:lifetime}), (\ref{eq:ineq_mc}) and (\ref{eq:ineq_cm}), respectively. The corresponding partial Lagrangian for problem (\ref{eq:CEopt}) can be expressed as 
\begin{eqnarray}\label{eq:lag}
&&\hspace{-0.7cm}\mathcal{L}(\{L_n^{\m \to \text{c}}\}, \{l_n^{\text{c}}\}, \{L_n^{\text{c} \to \m}\}, \mu, \{a_n\}, \{b_n\})\nonumber\\
&&\hspace{-0.7cm} = \sum_{n=1}^{N-2} E^{\m \to \text{c}}(L_n^{\m \to \text{c}}) + \mu \left(\sum_{n=1}^{N-2} E^{\text{c}}(l_{n+1}^{\text{c}}) + E^{\text{c} \to \m}(L_{n+2}^{\text{c} \to \m}) - E^{\text{c}}_0\right)\nonumber\\
&&\hspace{3.7cm} + \sum_{n=1}^{N-2}\sum_{i=1}^{n}a_n\left(l_{i+1}^{\text{c}} -L_{i}^{\m \to \text{c}}\right)+\sum_{n=1}^{N-2}\sum_{i=1}^{n}b_n\left(L_{i+2}^{\text{c} \to \m} - \kappa l_{i+1}^{\text{c}}\right),\nonumber\\
&&\hspace{-0.7cm} = \sum_{n=1}^{N-2} E^{\m \to \text{c}}(L_n^{\m \to \text{c}}) + \mu \left(\sum_{n=1}^{N-2} E^{\text{c}}(l_{n+1}^{\text{c}}) + E^{\text{c} \to \m}(L_{n+2}^{\text{c} \to \m}) - E^{\text{c}}_0\right)\nonumber\\
&&\hspace{3.7cm} - \sum_{n=1}^{N-2}\alpha_n L_{n}^{\m \to \text{c}} + \sum_{n=1}^{N-2}(\alpha_n - \kappa \beta_n)l_{n+1}^{\text{c}} + \sum_{n=1}^{N-2} \beta_nL_{n+2}^{\text{c} \to \m},\nonumber\\
\end{eqnarray} 
where we have defined $\alpha_n=\sum_{i=n}^{N-2}a_i$ and $\beta_n=\sum_{i=n}^{N-2}b_i$. It follows that the dual function for problem (\ref{eq:CEopt}) with respect to constraints (\ref{eq:eq_Lmc}) - (\ref{eq:eq_Lcm}) is given as
\begin{eqnarray}\label{eq:dualfunc}
&& \hspace{-0.7cm} g(\mu, \{a_n\}, \{b_n\})\\
&& \hspace{-0.7cm} = \hspace{-0.1cm} \left\{\begin{array}{llll}
\hspace{-0.3cm} \underset{\{L_n^{\m \to \text{c}}\}, \{l_n^{\text{c}}\}, \{L_n^{\text{c} \to \m}\}} \min & \hspace{-0.8cm} \mathcal{L}(\{L_n^{\m \to \text{c}}\}, \{l_n^{\text{c}}\}, \{L_n^{\text{c} \to \m}\}, \mu, \{a_n\}, \{b_n\}),\nonumber\\
\hspace{0.9cm}\text{s.t.} & \hspace{-0.8cm} \text{(\ref{eq:eq_Lmc}) - (\ref{eq:eq_Lcm})},\nonumber\\
\hspace{0.9cm} & \hspace{-0.8cm} L_n^{\m \to \text{c}} \ge 0,  l_{n+1}^{\text{c}} \ge 0, \hspace{0.1cm} \text{and} \hspace{0.1cm} L_{n+2}^{\text{c} \to \m} \ge 0, \hspace{0.1cm}  \text{for} \hspace{0.1cm} n=1, \dots, N-2,
\end{array}
\right.
\end{eqnarray}
and the dual problem is defined as 
\begin{equation}\label{eq:dual}
\underset{\mu, \{a_n\}, \{b_n\} \ge 0}{\text{maximize}} \hspace{0.3cm} g(\mu, \{a_n\}, \{b_n\}).
\end{equation}  
It is observed that, for any values of the Lagrange multipliers $(\mu, \{a_n\}, \{b_n\})$, the dual function $g(\mu, \{a_n\}, \{b_n\})$ can be decomposed as 
\begin{equation}
g(\mu, \{a_n\}, \{b_n\})=g^{\m \to \text{c}}(\{a_n\}) + g^\text{c}(\mu, \{a_n\}, \{b_n\}) + g^{\text{c} \to \m}(\mu, \{b_n\}),
\end{equation}
where we have defined the functions
\begin{subequations}\label{eq:sub}
\begin{eqnarray}
g^{\m \to \text{c}}(\{a_n\}) &=& \left\{\begin{array}{ll}
\hspace{-0.3cm}\underset{\{L_n^{\m \to \text{c}}\}} \min &\hspace{-0.3cm} \sum_{n=1}^{N-2} E^{\m \to \text{c}}(L_n^{\m \to \text{c}}) - \sum_{n=1}^{N-2}\alpha_n L_{n}^{\m \to \text{c}}, \\
\hspace{0cm}\text{s.t.} & \hspace{-0.3cm} \text{(\ref{eq:eq_Lmc}) and } L_n^{\m \to \text{c}} \ge 0, \hspace{0.1cm} \text{for} \hspace{0.1cm} n=1, \dots, N-2, 
\end{array}
\right. \label{eq:gmc}\\
g^\text{c}(\mu, \{a_n\}, \{b_n\}) &=& \left\{\begin{array}{ll}
\hspace{-0.2cm}\underset{\{l_n^{\text{c}}\}} \min & \hspace{-0.2cm} \mu\sum_{n=1}^{N-2} E^{\text{c}}(l_{n+1}^{\text{c}}) + \sum_{n=1}^{N-2}(\alpha_n - \kappa \beta_n)l_{n+1}^{\text{c}}, \\
\hspace{-0.1cm}\text{s.t.} &\hspace{-0.2cm} \text{(\ref{eq:eq_lc}) and } l_{n+1}^{\text{c}} \ge 0, \hspace{0.1cm} \text{for} \hspace{0.1cm} n=1, \dots, N-2,
\end{array}
\right. \label{eq:gc}\\
g^{\text{c} \to \m}(\mu, \{b_n\}) &=& \left\{\begin{array}{ll}
\hspace{-0.3cm}\underset{\{L_n^{\text{c} \to \m}\}} \min &\hspace{-0.3cm} \mu \sum_{n=1}^{N-2}E^{\text{c} \to \m}(L_{n+2}^{\text{c} \to \m})- \mu E^{\text{c}}_0 + \sum_{n=1}^{N-2} \beta_nL_{n+2}^{\text{c} \to \m}, \\
\hspace{-0.1cm}\text{s.t.} &\hspace{-0.3cm} \text{(\ref{eq:eq_Lcm}) and } L_{n+2}^{\text{c} \to \m} \ge 0, \hspace{0.1cm} \text{for} \hspace{0.1cm} n=1, \dots, N-2. \label{eq:gcm}
\end{array}
\right.
\end{eqnarray}
\end{subequations}

Based on the observations above, we tackle the original problem (\ref{eq:CEopt}) via its dual (\ref{eq:dual}) by means of the subgradient method over the multipliers $\mu$, $\{a_n\}$ and $\{b_n\}$ and by computing (\ref{eq:dualfunc}) via the solution of the three parallel subproblems (\ref{eq:gmc}), (\ref{eq:gc}) and (\ref{eq:gcm}). It is observed that, since the dual problem (\ref{eq:dual}) is strictly convex, the primal solution obtained at convergence is guaranteed to solve also the original problem (\ref{eq:CEopt}) [27]. The advantage of dual decomposition is that the three subproblems in (\ref{eq:sub}) are defined over a smaller domain with respect to the original problem and can be solved by imposing the Karush-Kuhn-Tucker (KKT) conditions. In fact, three subproblems are convex and satisfy the linearity constraint qualification since all the inequality and equality constraints are affine functions [27, Sec. 5.2]. Accordingly, as proved in Appendix \ref{app:opt_sol}, the respective solutions of problems (\ref{eq:gmc}), (\ref{eq:gc}) and (\ref{eq:gcm}) can be found as
\begin{subequations}\label{eq:opt_sol}
\begin{eqnarray}
L_n^{\m \to \text{c}, \opt}&=&\left[B\delta \log_2 \frac{h_n}{N_0 \ln 2}\left(\lambda + \alpha_n\right)\right]^+, \label{eq:sol_mc}\\
l_{n+1}^{\text{c}, \opt}&=&\sqrt{\frac{\delta^2}{3\mu\gamma^{\text{c}}C^3}\left[\nu - \alpha_n + \kappa \beta_n \right]^+}, \label{eq:sol_c}\\
L_{n+2}^{\text{c} \to \m, \opt}&=&\left[B\delta \log_2 \frac{h_{n+2}}{\mu N_0 \ln 2}\left(\eta - \beta_n\right)\right]^+,\label{eq:sol_cm}
\end{eqnarray}
\end{subequations}
for $n=1, \dots, N-2$, where $[x]^+=\max\{x,0\}$, and $\lambda$, $\nu$ and $\eta$ are parameters, each chosen so as to guarantee equality in the constraint (\ref{eq:eq_Lmc}), (\ref{eq:eq_lc}) or (\ref{eq:eq_Lcm}), respectively. Parameter $\lambda$, $\nu$ and $\eta$ can hence be computed separately via the standard bisection method [27]. 

The overall subgradient-based procedure is summarized in Algorithm \ref{al1}, where the subgradients of $g(\mu, \{a_n\}, \{b_n\})$ at point $(\mu, \{a_n\}, \{b_n\})$ are given as $(s^\mu, \{s_n^a\}, \{s_n^b\})$ with $s^\mu = (\sum_{n=1}^{N-2}$ $E_{n+1}^{\text{c}}(l_{n+1}^{\text{c}, \opt})+ E_{n+2}^{\text{c} \to \m}(L_{n+2}^{\text{c} \to \m, \opt})) - E^{\text{c}}_0$, $s_n^a= \sum_{i=1}^{n}(l_{i+1}^{\text{c}, \opt} -L_{i}^{\m \to \text{c}, \opt})$ and $s_n^b= \sum_{i=1}^n (L_{i+2}^{\text{c} \to \m,  \opt} - \kappa l_{i+1}^{\text{c}, \opt})$ for $n=1, \dots, N-2$. 
\begin{algorithm}[h]
\begin{algorithmic}
\caption{Optimal Bit Allocation} \label{al1} 
\State {\textbf{Initialization}}: $\mu \ge 0$, $\{a_n \ge 0\}$ and $\{b_n \ge 0\}$ for $n=1, \dots, N-2$.
\State  {\textbf{Repeat until convergence}}:  
\State \indent Obtain $\{L_n^{\m \to \text{c}, \opt}\}$, $\{l_{n+1}^{\text{c}, \opt}\}$ and $\{L_{n+2}^{\text{c} \to \m, \opt}\}$ using (\ref{eq:opt_sol}).
\State \indent Compute the subgradients of $g(\mu, \{a_n\}, \{b_n\})$.
\State \indent Update $\mu$, $\{a_n\}$ and $\{b_n\}$ using the subgradient method.
\State {\textbf{Output:}}  $\{L_n^{\m \to \text{c}, \opt}\}$, $\{l_{n+1}^{\text{c}, \opt}\}$ and $\{L_{n+2}^{\text{c} \to \m, \opt}\}$ for $n=1, \dots, N-2$.
\end{algorithmic}
\end{algorithm}
\section{Numerical Results}\label{sec:num} 
\begin{table}[t]
\caption{Simulation Parameters} \label{t1}
\begin{center}
    \begin{tabular}{|p{2.5cm}|p{9cm}|p{3.5cm}|}
    \hline
    Parameter & Definition & Value \\ \hline\hline
    $B$ & Bandwidth & $20$ MHz \\ \hline
    $N_0$ & Noise spectrum density & $-174$ dBm/Hz \\ \hline
    $h_0/(N_0B)$ & Reference SNR & $20$ dB \\ \hline
    $C$ & Number of CPU cycles per bit ($95$th percentile of random number of cycles in [22], [23]) & $1550.7$\\ \hline
    $\gamma^\m$ & Switch capacitance constant of mobile & $10^{-28}$ [22], [23] \\ \hline
    $\gamma^{\text{c}}$ & Switch capacitance constant of cloudlet & $10^{-28}$ [22], [23] \\ \hline
    $L$ & Number of input bits & $15$ Mbits \\ \hline
    $\kappa$ & Number of output bits per input bits & $0.9$ \\ \hline
    $E_0^{\text{c}}$ & Energy budget of the UAV for given user & $100$ kJ\\ \hline
    $\delta$ & Slot duration & $2.5$ ms\\ \hline
    $\Delta$ & Frame duration & $100$ ms\\ \hline
    $\pmb{p}_0^{\text{c}}$ & UAV's initial position & $(5, 5, 5)$ m\\ 
    \hline
    \end{tabular}
    \end{center}
\end{table} 

\begin{figure}[t]
\begin{center}
\includegraphics[width=13cm]{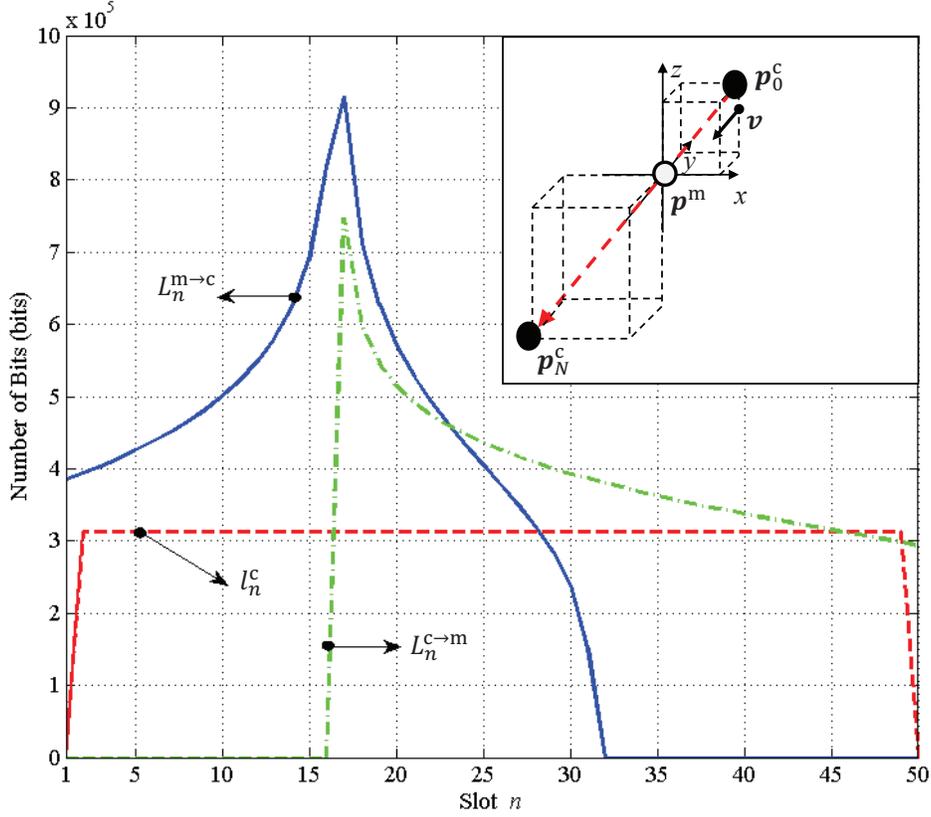}
\caption{Optimal bit allocation for the UAV trajectory indicated in the inset ($L = 15$ Mbits, $T=5$ s, $\delta = 2.5$ ms, $\Delta = 100$ ms, $E_0^{\text{c}}=100$ KJ, $\pmb{p}_0^{\text{c}}=(5, 5, 5)$ (m), $\pmb{p}_N^{\text{c}}=(-10, -10, -10)$ (m) and $\pmb{v}=(-3, -3, -3)$ (m/s)).}  \label{fig:traj}
\end{center}
\end{figure}

In this section, the performance of mobile cloud computing system based on a mobile cloudlet is investigated by means of numerical simulations. The focus is on comparing the performance of the optimal bit allocation scheme in Algorithm \ref{al1} with an equal bit allocation scheme in which $L_n^{\m \to \text{c}}=l_{n+1}^{\text{c}}=L/(N-2)$ and $L_{n+2}^{\text{c} \to \m}=\kappa L/(N-2)$ are set for $n=1, \dots, N-2$. The parameters are set as follows unless specified otherwise. The communication bandwidth per link is $B=20$ MHz, and the noise spectrum density is $N_0=-174$ dBm/Hz. The reference SNR $h_0/(N_0B)$ at distance $d_0 = 1$ m is assumed to be $20$ dB. In addition, the number of CPU cycles per bit is $C = 1550.7$, which corresponds to the $95$th-percentile of the random number of cycles used in [22], [23]. The switch capacitance constants of mobile and cloudlet are $\gamma^\m = \gamma^{\text{c}} = 10^{-28}$ [22], [23]. The number of input bits is set to be $L=15$ Mbits and the number of output bits per input bit is $\kappa = 0.9$. The available energy of the cloudlet is set for the given user as $E_0^{\text{c}} = 100$ KJ. Also, the slot duration and frame duration are chosen as $\delta = 2.5$ ms and $\Delta = 100$ ms. The UAV trajectory indicated in the inset of Fig. \ref{fig:traj} is considered, where the UAV starts at position $\pmb{p}_0^{\text{c}}=(5, 5, 5)$ (m) and flights unidirectionally towards the mobile device with velocity vector $\pmb{v}$ so that $\pmb{p}_n^{\text{c}}=\pmb{p}_0^{\text{c}} + n\pmb{v}\Delta$ for $n=1, \dots, N$. The above parameters are summarized in Table \ref{t1}.

First, the optimal bit allocations $\{L_n^{\m \to \text{c}}\}$, $\{l_c^{\text{c}}\}$ and $\{L_n^{\text{c}\to \m}\}$ obtained by Algorithm \ref{al1} are illustrated as a function of the slot index $n$ under the maximum latency constraint $T=5$ s with  UAV's velocity $\pmb{v}=(-3, -3, -3)$ (m/s). As shown in Fig. \ref{fig:traj}, a larger number $\{L_n^{\m \to \text{c}}\}$ of bits is allocated for uplink transmission when the UAV is closer to the mobile device. Nevertheless, in order to reduce the energy consumption at the UAV, it is preferable to process an equal number of bits in each slot. As a result, the mobile transmits to the UAV also when the UAV is not in the position closest to the mobile. Moreover, the bit allocation $\{L_n^{\text{c}\to \m}\}$ for downlink transmission depends not only on the position of the UAV, but also on the availability of the cloudlet output as a result of computing.  

\begin{figure}[t]
\begin{center}
\includegraphics[width=13cm]{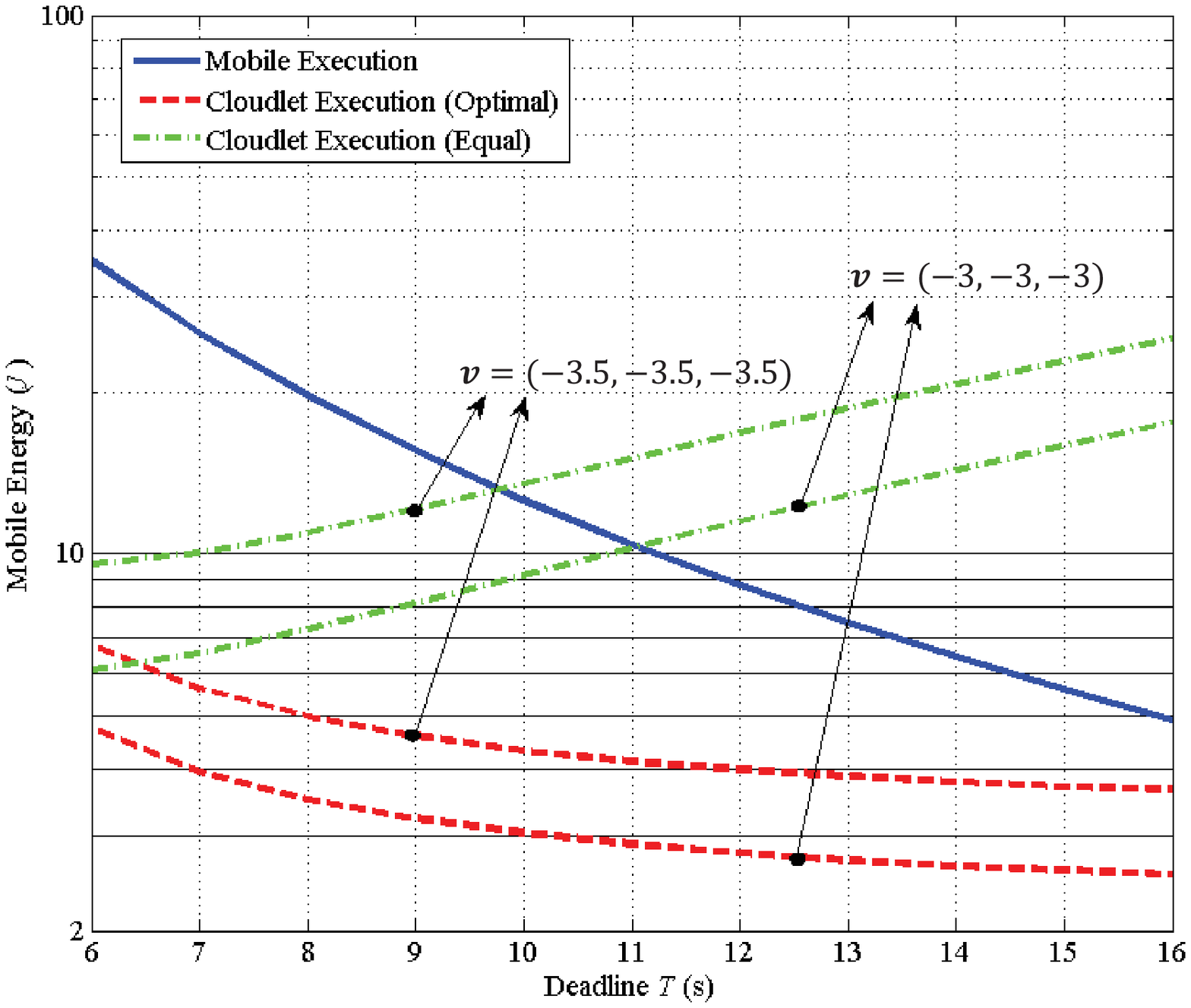}
\caption{Mobile energy consumption as a function of the deadline $T$ ($L = 15$ Mbits, $\delta = 2.5$ms, $\Delta = 100$ ms, $E_0^{\text{c}}=100$ KJ, $\pmb{p}_0^{\text{c}}=(5, 5, 5)$ (m) with $\pmb{v}= (-3, -3, -3)$ (m/s) and $\pmb{v}= (-3.5, -3.5, -3.5)$ (m/s)).}  \label{fig:T}
\end{center}
\end{figure}  

Then, the minimum mobile energy consumptions with mobile and cloudlet execution are compared in Fig. \ref{fig:T}, that is, $E^{\m}$ and $\sum_{n=1}^{N-2}E^{\m \to \text{c}}(L_n^{\m \to \text{c}, \opt})$ in (\ref{eq:optE_ME_tot}) and (\ref{eq:CEopt}), respectively, as a function of the deadline $T$ within which the input bits $L$ need to be processed with two different cloudlet's velocity vectors $\pmb{v}= (-3, -3, -3)$ (m/s) and $\pmb{v}= (-3.5, -3.5, -3.5)$ (m/s). It is first observed that optimal bit allocation significantly reduces energy consumption at the mobile device, particularly as the latency constraint $T$ increases. In fact, an equal bit allocation may even entail an increasing mobile energy consumption with $T$, as it forces communication in slots in which the UAV is far from the mobile device. When the deadline is stringent, cloudlet execution is seen to be more energy efficient than mobile execution, especially if the velocity vector $\pmb{v}$ is small, which ensures that the UAV will remain in the vicinity of the mobile for a large number of slots given the selected initial position. Additionally, it can be expected that the large workload $L$ has similar impact on the performance with the stringent $T$, in that the cloudlet execution becomes more efficient compared to the mobile execution. 

\section{Concluding Remarks}\label{sec:con} 
In this paper, a mobile cloud computing architecture is studied based on a UAV-mounted cloudlet that provides offloading opportunities to mobile devices in the absence of a dense infrastructure of base stations. Use cases include the support of rescue or military operations via image or video recognition software run on mobile devices for the assessment of the status of victims, enemies, or hazardous terrain and structures. The optimization of the offloading process for a static mobile device is studied with respect to the criterion of minimum mobile energy consumption. Numerical results validate the significant advantages of the proposed approach as a function of the UAV's trajectory. Interesting open problems concern the generalization of the optimization studied here to multiple static or moving interfering mobile devices with the UAV's path planning.
\section{Appendix}
\subsection{Derivations of (\ref{eq:opt_sol})}\label{app:opt_sol}
In this appendix, the optimal solutions are derived for the three parallel subproblems (\ref{eq:gmc}), (\ref{eq:gc}) and (\ref{eq:gcm}) by applying the KKT conditions. The Lagrangian functions associated to problem (\ref{eq:gmc}), (\ref{eq:gc}) and (\ref{eq:gcm}) are given as
\begin{eqnarray*}
\mathcal{L}^{\m \to \text{c}}(\{L_n^{\m \to \text{c}}\}, \{a_n\}, \lambda) &=& \sum_{n=1}^{N-2}\left(2^{\frac{L_n^{\m \to \text{c}}}{B\delta}}-1\right)\frac{N_0 B \delta}{h_n}  -\sum_{n=1}^{N-2} \alpha_nL_n^{\m \to \text{c}} + \lambda \left(L-\sum_{n=1}^{N-2}L_n^{\m \to \text{c}}\right)\\
\mathcal{L}^\text{c}(\{l_n^{\text{c}}\}, \mu, \{a_n\}, \{b_n\}, \nu) &=& \mu \sum_{n=1}^{N-2}\frac{\gamma^\text{c} C^3}{\delta^2}(l_{n+1}^{\text{c}})^3 + \sum_{n=1}^{N-2}\left(\alpha_n - \kappa \beta_n\right)l_{n+1}^{\text{c}} +\nu\left(L-\sum_{n=1}^{N-2} l_{n+1}^{\text{c}}\right) \\
\mathcal{L}^{\text{c} \to \m}(\{L_n^{\text{c}\to \m}\}, \mu, \{b_n\}, \eta) &=&  \mu \sum_{n=1}^{N-2}\left(2^{\frac{L_{n+2}^{\text{c} \to \m}}{B\delta}}-1\right)\frac{N_0 B \delta}{h_{n+2}}  - \mu E_0^\text{c} + \sum_{n=1}^{N-2}\beta_nL_{n+2}^{\text{c} \to \m} \nonumber\\
&+&\eta\left(\kappa L-\sum_{n=1}^{N-2} L_{n+2}^{\text{c} \to \m}\right) 
\end{eqnarray*} 
respectively. Then, the KKT conditions for (\ref{eq:gmc}), (\ref{eq:gc}) and (\ref{eq:gcm}) can be obtained as
\begin{eqnarray*}
\frac{\partial \mathcal{L}^{\m \to \text{c}}(\{L_n^{\m \to \text{c}}\}, \{a_n\}, \lambda)}{\partial L_n^{\m \to \text{c}}} &=& \frac{N_0\ln 2}{h_n}2^{\frac{L_n^{\m \to \text{c}}}{B\delta}} - \alpha_n - \lambda  = 0,\\
\frac{\partial \mathcal{L}^\text{c}(\{l_n^{\text{c}}\}, \mu, \{a_n\}, \{b_n\}, \nu)}{\partial l_{n+1}^{\text{c}}} &=& \frac{3\mu\gamma^{\text{c}}C^3}{\delta^2}(l_{n+1}^{\text{c}})^2 +\alpha_n-\kappa \beta_n -\nu = 0,\\
\frac{\partial \mathcal{L}^{\text{c} \to \m}(\{L_n^{\text{c}\to \m}\}, \mu, \{b_n\}, \eta)}{\partial L_{n+2}^{\text{c} \to \m}}&=& \frac{\mu N_0\ln 2}{h_{n+2}}2^{\frac{L_{n+2}^{\text{c} \to \m}}{B\delta}}+\beta_n -\eta= 0,
\end{eqnarray*}
for $n=1, \dots, N-2$, from which we can get the optimal solutions as in (\ref{eq:opt_sol}). 
{\parindent0pt
\parskip8pt
\section{References}
[1] Dinh, H., Lee, C., Niyato, D.,~\textit{et al}.: 'A survey of mobile cloud computing: architecture, applications, and approaches', Wireless communications and mobile computing, 2013, \textbf{13}, (18), pp. 1587--1611

[2] Satyanarayanan, M., Bahl, R. C. P., Davies, N.: 'The case for VM-based cloudlets in mobile computing', IEEE Pervasive Computing, 2009, \textbf{8}, (4), pp. 14--23

[3] Frew, E. W., Brown, T. X.: 'Airborne communication networks for small unmanned aircraft systems', Proceedings of the IEEE, 2008, \textbf{96}, (12), pp. 2008--2027

[4] Zeng, Y., Zhang, R., Lim, T. J.: 'Wireless communications with unmanned aerial vehicles: Opportunities and challenges',  IEEE Communications Magazine, 2016, \textbf{54}, (5), pp. 36--42

[5] Orfanus, D., De Freitas, E. P., Eliassen, F.: 'Self-organization as a supporting paradigm for military UAV relay networks', IEEE Communications Letters, 2016, \textbf{20}, (4), pp. 804--807

[6] Loke, S. W.: 'The Internet of Flying-Things: Opportunities and challenges with airborne fog computing and mobile cloud in the clouds', arXiv preprint arXiv:1507.04492, 2015

[7] Bonomi, F., Milito, R., Natarajan, P.,~\textit{et al}.: 'Fog computing: A platform for Internet of Things and analytics', 'Big data and Internet of Things: A roadmap for smart environments' (Springer International Publishing, 2014), pp. 169--186

[8] Zeng, Y., Zhang, R., Lim, T.J.: 'Throughput maximization for UAV-enabled mobile relaying systems', IEEE Trans. Commun., 2016, \textbf{64}, (12), pp. 4983–4996

[9] Zhao, W., Ammar, M., Zegura, E.: 'A message ferrying approach for data delivery in sparse mobile ad hoc networks'. Proc. ACM International Symposium on Mobile Ad Hoc Networking and Computing, Tokyo, Japan, May 2004, pp. 187--198

[10] Shah, R., Roy,  S., Jain, S.,~\textit{et al}.: 'Data MULEs: Modeling and analysis of a three-tier architecture for sparse sensor networks', Ad Hoc Networks, 2003, \textbf{1}, (2), pp. 215--233

[11] Zhan, P., Yu, K., Swindlehurst, A. L.: 'Wireless relay communications with unmanned aerial vehicles: Performance and optimization', IEEE Trans. Aerospace and Electronic Systems, 2011, \textbf{47}, (3), pp. 2068--2085

[12] Li, J., Han, Y.: 'Optimal resource allocation for packet delay minimization in multi-layer UAV networks', IEEE Commun. Lett., 2017, \textbf{21}, (3), pp. 580–583

[13] Soorki, M. N., Mozaffari, M., Saad, W.: 'Resource allocation for machine-to-machine communications with unmanned aerial vehicles'. Proc. IEEE Global Communications Conference (GLOBECOM), Washington, DC, USA, Dec. 2016

[14] Mozaffari, M., Saad, W., Bennis, M.,~\textit{et al}.: 'Mobile Internet of Things: Can UAVs provide an
energy-efficient mobile architecture?'. Proc. IEEE Global Communications Conference (GLOBECOM), Washington, DC, USA, Dec. 2016

[15] Mozaffari, M., Saad, W., Bennis, M.,~\textit{et al}.: 'Drone small cells in the clouds: Design, deployment and performance analysis'. Proc. IEEE Global Communications Conference (GLOBECOM), San Diego, CA, USA, Dec. 2015

[16] Mozaffari, M., Saad, W., Bennis, M.,~\textit{et al}.: 'Optimal transport theory for power-efficient deployment of unmanned aerial vehicles'. Proc. IEEE International Conference on Communications (ICC), Kuala Lumpur, Malaysia, May 2016

[17] Mozaffari, M., Saad, W., Bennis, M.,~\textit{et al}.: 'Unmanned aerial vehicle with underlaid Device-to-Device communications: Performance and tradeoffs', IEEE Trans. on Wireless Communications, 2016, \textbf{15}, (6), pp. 3949--3963

[18] Lyu, J., Zeng, Y., Zhang, R.,~\textit{et al}.: 'Placement optimization of UAV-mounted mobile base stations', IEEE Commun. Lett., 2017, \textbf{21}, (3), pp. 604–607

[19] Zeng, Y., Zhang, R.: 'Energy-efficient UAV communication with trajectory optimization', arXiv preprint arXiv:1608.01828v1, 2016

[20] Borst, C., Sjer, F., Mulder,  M.,~\textit{et al}.: 'Ecological approach to support pilot terrain awareness after total engine failure', Journal of Aircraft, 2008, \textbf{45}, (1), pp. 159--171

[21] Chakrabarty, A., Langelaan, J.: 'Energy-based long-range path planning for soaring-capable unmanned aerial vehicles', Journal of Guidance, Control, and Dynamics, 2011, \textbf{34}, (4), pp. 1002--1015

[22] Yuan, W. H., Nahrstedt, K.: 'Energy-efficient soft real-time CPU scheduling for mobile multimedia systems', ACM SIGOPS Operating Systems Review, 2003, \textbf{37}, (5), pp. 149--163

[23] Yuan, W. H., Nahrstedt, K.: 'Energy-efficient CPU scheduling for multimedia applications', ACM Transactions on Computer Systems, 2006, \textbf{24}, (3), pp. 292--331

[24] Burd, T. D., Brodersen, R. W.: 'Processor design for portable systems', in 'Technologies for wireless computing' (Springer US, 1996), pp. 119--137

[25] Cover, T. M., Thomas, J. A.: 'Element of information theory' (John Wiley \& Sons, 2012)

[26] Palomar, D., Chiang, M.: 'A tutorial on decomposition methods for network utility maximization', IEEE Journal on Selected Areas in Communications, 2006, \textbf{24}, (8), pp. 1439--1451

[27] Boyd, S. P., Vandenberghe, L.: 'Convex optimization' (Cambridge University Press, 2004)

\end{document}